\documentstyle[11pt,newpasp_aod,twoside,epsf]{article}
\newlength{\zwidth}
\newcommand{\as}{\mbox{$''$}}
\newcommand{\ebv}{\mbox{$E(B-V)$}}
\newcommand{\etal}{{et~al.}}
\def\gapeq{\mathrel{\hbox{\rlap{\hbox{\lower4pt\hbox{$\sim$}}}\hbox{$>$}}}}
\def\lapeq{\mathrel{\hbox{\rlap{\hbox{\lower4pt\hbox{$\sim$}}}\hbox{$<$}}}}
\newcommand{\Lya}{\mbox{Ly{\footnotesize$\alpha$}}}
\newcommand{\zsp}{\mbox{\hspace{\zwidth}}}

\def\edcomment#1{\iffalse\marginpar{\raggedright\sl#1\/}\else\relax\fi}
\reversemarginpar

\begin{document}
\title{Dust in the High Redshift Universe }
 \author{G.R. Meurer}
\affil{Department of Physiscs and Astronomy, The Johns Hopkins
 University, Baltimore, MD 21218}

\begin{abstract}
This paper reviews the dust content of the high redshift ($z > 2$)
universe.  Studies of the various ``species'' in the high-$z$ ``zoo''
show that almost all have strong evidence for containing dust.  The one
exception, where the evidence is not yet convincing, is in quasar
absorption line systems, particularly those with low column density.
These may not even be associated with galaxies.  The high-$z$ galaxy
types which do show evidence for dust are all strongly star forming.
Hence, as seen locally, star formation and dust in the distant universe
are also strongly correlated.  It is beyond debate that star formation
at $z \sim 3$ is dominated by dusty systems who emit most of their
bolometric flux in the rest-frame FIR.  What is not clear is whether
these systems are totally invisible at shorter wavelengths, or whether a
large fraction are visible in the rest-frame UV as Lyman Break Galaxies.
The issue may not be settled until the sub-mm background is definitively
resolved. 
\end{abstract}

\section{Introduction}

Here I review some of the things we know about dust in the high redshift
($z \gapeq 2$) universe.  This review is meant to be an introduction to
the literature.  While by necessity incomplete, it is my hope that the
reader will get an appreciation for some of the activity of the field
and the researchers involved.

To many extraglactic astronomers and cosmologists dust is at best a
nuisance.  However, it is fairly ubiquitous and perilous to ignore.  For
example, Blakeslee \etal\ (2003), following on from Aguirre (1999) and
Goobar \etal\ (2002), show that the
distinctive signature of cosmological acceleration in the differential
Hubble diagram can also be mimicked with a model containing gray dust
having fixed spatial density in the intergalactic medium (IGM).  While
there are a variety of reasons to believe that the effects are not
significant (e.g.\ Aguirre \&\ Haimann 2000; Paerels \etal\ 2002; also
cf.\ \S2.4 below), it is too early to totally rule out the gray dust
model.

As in the local universe, dust is highly correlated with star formation.
Since all studies show that the co-moving star formation rate density
(SFRD) monotonically increases with $z$ out to a redshift of at least 1
(e.g.\ Madau \etal\ 1996) we expect that dust may be increasingly
important at higher redshift (Calzetti \&\ Heckman 1999).  Dust is
especially important in interpreting the cosmic star formation history
as shown in so called ``Madau Plots'' of SFRD$(z)$. In some early
studies dust was ignored leading to a view that SFRD$(z)$ peaked around
$z \sim 1$ and declined towards higher redshift (e.g.\ Madau \etal\
1996). It is now recognized that most of the UV emission from high-$z$
star formation is at least somewhat obscured by dust (e.g.\ Madau,
Pozzetti, \&\ Dickinson 1998; Pettini \etal\ 1998) leading to SFRD$(z)$
plots that flatten for $z \gapeq 1$ (e.g.\ Calzetti 1999).

Since observations of high-$z$ star forming galaxies are somewhat easier
in the rest-frame UV than in the rest-frame far-infrared (FIR), it is of
interest to know if star formation rates can be recovered from
rest-frame UV observations through suitable application of reddening
laws.  More importantly we must ask whether the numerous high-redshift
galaxies detected in the rest frame UV are really significant in terms
of the total star formation happening in the early universe.

This review is broken into two main sections.  \S~2 reviews dust content of
the species in the ``high-$z$ zoo''.  \S~3 reviews the evidence for
galaxy scale reddening laws working at $z > 2$ and considers the debate
as to whether the majority of star formation is totally hidden from the
rest-frame UV and optical at high redshift.  \S~4 briefly summarizes
breaking news and the expected progress in the field.

\section{Evidence for High Redshift Dust\label{s:zoo}}

High-redshift objects can be selected by many ways at a variety of
wavelengths.  While there is some overlap in the properties of the
galaxies selected by the different techniques they typically go by
different names in the literature.  Table~1 summarizes the
beasts in the high-$z$ zoo.  

\begin{table}
\caption{The high-$z$ zoo}
\begin{tabular}{lll}&& \\[-2mm] \hline \hline \\[-3mm]
\multicolumn{2}{c}{wavelength} & \\
observed & rest & name \\ &&\\[-3mm] \hline \\[-3mm]
optical  & UV   & High-$z$ quasars \\
         &      & Lyman Break Galaxies (LBG) \\
         &      & Lyman-$\alpha$ galaxies \\
         &      & Quasar absorption line systems \\ &&\\[-3mm] \hline \\[-3mm]
NIR & optical   & Extremely Red Objects (ERO) \\ &&\\[-3mm] \hline \\[-3mm]
sub-mm/mm & FIR & Sub-mm and mm galaxies, SCUBA galaxies \\ &&\\[-3mm] \hline \\[-3mm]
radio    & radio& Micro-Jansky Radio Sources \\ &&\\[-3mm] \hline \\[-3mm]
\end{tabular}
\end{table}

{\bf 2.1. High-$z$ quasars}.  Due to their high-luminosities, quasars are
relatively easy to detect out to the highest known redshifts (e.g.\
Irwin \&\ McMahon, 1991; Fan \etal\ 2001; 2003).  The current record
holder has $z = 6.43$ (Fan \etal 2003).  One of the most early and
direct measurements of high-$z$ dust came from Omont \etal\ (1996) who
observed the $z = 4.7$ radio quiet quasar BR1202-0725 with the IRAM
array. They simultaneously detected both mm wavelength dust continuum
emission as well as molecular CO (5-4) and (7-6) lines in emission.  The
mm observations show two sources: the quasar and a companion galaxy.  The
dust emission of the latter is certainly powered by star-formation.  So
in this system we see star-formation and it's key ingredients, dust and
molecular gas.

{\bf 2.2. Lyman Break Galaxies}.  Lyman Break Galaxies are selected
using broad band filters in the optical and/or NIR.  The strong spectral
break at $\lambda = 912$\AA, due to the ionization of hydrogen, is
readily seen in broad-band SEDs.  For example, galaxies with $z \sim 2.8$
will have no flux in the HST/WFPC2 F300W ($U$) band but can be detected
at longer wavelengths, hence the term $U$-dropout (or $B$-dropout,
$V$-dropout, etc.).  LBGs are typically selected for having a very red
color (or color limit) using filters that straddle the break, but
relatively blue color using filters that are longwards of the Lyman
break (e.g.\ Madau \etal\ 1996; Steidel \etal\ 2003).  This ensures the
selection of star forming galaxies, and selects against very old or very
dust reddened galaxies.  The ability to obtain optical photometry to the
25th magnitude and beyond with current technology means that LBGs are
the most abundant species in the High-$z$ zoo (Adelberger \&\ Steidel
2000, hereafter AS00).

LBGs display a remarkable similarity to nearby UV bright starburst
galaxies as defined by the IUE atlas of Kinney \etal\ (1993).  Both
types have rest-frame SEDs that are at the blue end of those found for
normal galaxies (Papovich \etal\ 2001); strong emission lines in the
rest-frame optical (Pettini \etal\ 1998); rest-frame UV spectra
dominated by high ionization wind lines as well as strong narrow
interstellar absorption lines (Tremonti \etal\ 2001; Shapley \etal\
2003); and a net blue shift of the interstellar absorption lines with
respect to photospheric stellar lines, indicative of strong outflows in
the ISM (Pettini \etal\ 2000; Shapley \etal\ 2003).  The main difference
is that LBGs are much more luminous (e.g.\ AS00), typically by an order
of magnitude or more even without any dust corrections.  Since the two
types have similar high effective surface brightnesses (Meurer \etal\
1997), LBGs are also much larger, typically having effective radii of a
few kpc (Giavalisco \etal\ 1996).  In short, LBGs look like local UV
bright starbursts but scaled up in size and hence luminosity.

While we know a lot about the dust content of the local UV bright
starburst population (e.g.\ Calzetti \etal\ 1994; 2000; Calzetti 2001;
Meurer \etal\ 1995; Gordon, Calzetti \&\ Witt 1997; Meurer \etal\ 1999, 
hereafter MHC99), much less is known directly about the dust content of
LBGs.  Most individual LBGs are not detected in the sub-mm with the
Submillimeter Common User Bolometer Array (SCUBA), although there are a
few rare exceptions (AS00, Baker \etal\ 2001). Stacked SCUBA studies
also have had mixed success in detecting LBGs (AS00; Chapman \etal\
2000; Peacock \etal\ 2000).  Nevertheless, we can infer the presence of
dust in LBGs via reddening: we know that they have a strong ionizing
population from their emission line spectrum, yet their colors are not
as blue as expected from un-reddened stellar populations.  Using
broad-band SEDs Papovich \etal\ (2001) estimate the reddening
distribution of LBGs which peaks at $\ebv \approx 0.15$.  A variety of
studies starting with Meurer \etal\ (1997) have estimated typical UV
attenuations in LBG samples using just rest-frame UV colors.  Most
recent studies estimate UV attenuation factors around 5 using this
method.  One of the most recent works in this vein is Vijh, Witt \&\
Gordon (2003) who also present a nice compilation of attenuation
estimates for LBGs.  This subject is addressed in more detail in \S{3}.

{\bf 2.3. Lyman-$\alpha$ galaxies}.  Initially \Lya\ emission was
thought to be one of the best ways to detect the first epoch of galaxy
formation (Partridge \&\ Peebles 1967).  However, after many
disappointing surveys it was realized that there was something wrong
with the original predictions (e.g.\ Pritchet 1994).  Spectroscopic
follow-up of LBGs showed that they have low rest-frame \Lya\ equivalent
widths ($\lapeq 20$\AA) and fluxes lower than expected for their UV
continuum strength (e.g. Steidel \etal\ 1996a,b; Lowenthal \etal\ 1997).
This is due to the effects of resonant scattering of the \Lya\ photons
through the ISM of the galaxies, which greatly increases the total path
required for the photons to escape the system.  Hence even a small
amount of dust is enough to greatly attenuate \Lya\ emission compared to
the neighboring continuum.  In an expanding dusty ISM, such as a
galactic wind, \Lya\ photons can escape by back scattering out the
far-side of the outflow resulting in a distinctly asymmetric line
profile characterized by a sharp blue side cutoff.  Such \Lya\ profiles
are indeed observed in both nearby starbursts (Kunth \etal\ 1998) as
well as LBGs (Shapley \etal\ 2003).

Recent high-$z$ \Lya\ surveyors have learned their lessons and are going
deeper and wider, and hence are becoming more successful (e.g.\ Rhoads
\etal\ 2000).  In fact the current most distant ``normal'' galaxies ($z
\sim 6.5$) have been found using narrow band imaging targeting \Lya\
(Kodaira \etal\ 2003). Spectroscopic confirmation of these and other
blank-field \Lya\ emitters inevitably shows the asymmetric profiles
indicating the presence of a dusty expanding ISM (Kodaira \etal\ 2003;
Rhoads \etal\ 2000; 2003).

{\bf 2.4. Quasar Absorption Line Systems}.  The absorption lines in the
spectra of quasars probe the gas phase of intervening systems, which
need not necessarily be galaxies (self-gravitating
conglomerations of stars, gas and dark matter).  The most commonly
observed feature seen is \Lya\ absorption.  In the literature, \Lya\
absorption lines are referred to (in order of decreasing $\log(N_{\rm
HI})$) as ``damped \Lya\ absorption systems'' (DLAS) with $\log(N_{\rm
HI} [{\rm atoms\, cm^{-2}}]) \gapeq 20$, ``Lyman limit systems'' with
$17 \gapeq \log(N_{\rm HI}) \gapeq 20$, and ``Lyman forest'' clouds $14
\gapeq \log(N_{\rm HI}) \gapeq 17$.

Metal absorption lines are also seen, albeit much less frequently.
These include C, Si, Mg, S, Zn.  Metals are also seen in the
IGM out to $z \sim 5$, with little evolution in the cosmic density in
\ion{C}{iv} absorption for $1.5 \lapeq z \lapeq 5.5$ (Songaila 2001;
Pettini \etal\ 2003).  This lack of evolution is somewhat puzzling.  It
may indicate a massive pollution event to the IGM at $z > 5.5$ (Songaila
2001), or alternatively may be due to IGM features being correlated with
star formation which also evolves only weakly with redshift (Adelberger
\etal\ 2003).

It is less clear that the IGM contains significant quantities of dust.
We expect a bias against detecting dusty IGM clouds in optically
selected quasar samples - the dust would diminish the flux of the
background quasar (Fall \&\ Pei, 1993).  However, a study of a radio
selected sample of quasars shows that the dust bias is at most a factor
of two in absorption line systems having $2 \lapeq z \lapeq 3$
(Ellingson \etal\ 2001).  Prochaska \etal\ (2003) provide some evidence
suggestive of dust in a DLAS at $z = 2.6$: the elemental abundance
pattern of this system scales well to the solar abundance {\em after
correction for depletion onto dust grains}.  This on its own is not
convincing proof of dust in DLASs, nor does it follow that lower column
density sight lines of the IGM contain dust.

{\bf 2.5. Extremely Red Objects}.  Elston \etal\ (1988) pointed out the
existence of an interesting new population of galaxies having very red
optical - NIR colors, $R - K \gapeq 5$.  These ``Extremely Red Objects''
have cropped up in numerous other deep NIR surveys although the exact
selection limits vary.  Detailed studies of the multi-wavelength SEDs of
EROs show that they are a mixed bag, with roughly half being dusty
starbursts and the other half being passively evolving (presumably
dust-free) ellipticals at $z \sim 1$ (e.g.\ Smail \etal\ 2002a).
Likewise, Ivison \etal\ (2002) find that about half of the
radio-confirmed bright SCUBA sources have ERO or very red counterparts
showing the strong overlap between the SCUBA and ERO populations.

By selecting purely in the NIR it is possible to select the most extreme
galaxies - the Hyper Extremely Red Objects or HEROs (Totani \etal\ 2001)
with colors $J - K \gapeq 3$ so red that they can not be produced by
pure passive evolution - some dust is required.  Totani \etal\ find that
the these are best modeled as very dusty starbursts at $z \sim 3$.

{\bf 2.6. Sub-mm/mm Galaxies}.  The advent of SCUBA on the 15m James Clerk
Maxwell Telescope made it possible to survey for
dust emission from high-luminosity, high-$z$ galaxies.  SCUBA has been
particularly effective at 850\micron, where ``negative $K$-corrections''
result in sources with fixed star formation rate having nearly constant
flux as a function of $z$ in the range of $\sim 0.5$ to 5 (Guiderdoni
\etal\ 1997). The 850\micron\ confusion limit for SCUBA is $\sim$2 mJy
(Hughes \etal\ 1998) corresponding to Bolometric luminosities of $\sim 2
\times 10^{12}\, L_\odot$, about that of Arp 220.  Hence only
ultra-luminous ($L_{\rm bol} > 10^{12}\, L_\odot$) and hyper-luminous
galaxies ($L_{\rm bol} > 10^{13}\, L_\odot$) are detectable with SCUBA
in blank fields.  Similar star formation rate detection levels are also
possible at 1.2mm, with using the MAMBO detector on the 30m IRAM
telescope (Dannerbauer \etal\ 2002). Staring at strong lensing clusters
allows the detection limit to be pushed down by a typical factor of
$\sim 3$ (Smail \etal\ 2002b).  The resulting lens amplification
corrected number counts indicates that the 850\micron\ background is
nearly completely resolved at sub-mJy levels.

The large beam sizes of SCUBA (15\as) and MAMBO (11\as) make
identification of optical counterparts difficult.  The optical
counterparts are usually faint, and often not the most obvious galaxy in
the sub-mm beam; Frayer \etal\ (2003) show an example of such an
identification.  Radio synthesis follow-up studies allows the
counterparts of sub-mm and mm galaxies to be pinpointed to sub-arcsec
accuracy.  Ivison \etal\ (2002) find that 60\%\ of their 850\micron\ ``8
mJy sample'' have robust radio identifications, and that 90\%\ of those
identified in the radio have near-infrared (NIR) and optical
counterparts.  Hence, over half of the brightest SCUBA sources have rest
frame UV and optical counterparts.  The success rate for finding optical
and NIR counterparts for fainter highly magnified lensed SCUBA sources
appears to be lower (Smail \etal\ 2002b), although the radio detection
limits tend not to be that deep in those cases.  The high dust luminosity and
faint rest-frame UV and optical fluxes indicate that sub-mm/mm galaxies
are similar to local Ultra Luminous Infrared Galaxies (ULIRGs; e.g.\
Goldader \etal\ 2002), but scaled up in luminosity.

Details and further information on sub-mm galaxies can be found in the
excellent and comprehensive review of Blain \etal\ (2002).

{\bf 2.7. Micro-Jansky Radio Sources}.  The radio emission of local star
forming galaxies correlates very well with the FIR emission, although
the physics behind the correlation is less clear (Helou \etal\ 1985; de
Jong \etal\ 1985; Lisenfeld \etal\ 1996). Hence, radio observations are
an excellent means to peer through the dust in a galaxy and observe star
formation.  For sources with very low fluxes, $f_\nu(3.5{\rm cm}) \lapeq
35\, \mu$Jy, the frequency domain spectral slope $\alpha$ ($f_\nu
\propto \nu^\alpha$`) becomes more steep ($\alpha ~ -0.7$) indicating
that star forming galaxies are dominating over AGN (Fomalont \etal\
2002).  In contrast to the sub-mm, star forming galaxies dim
considerably with redshift.  Hence, spectroscopic, follow-up studies
show that the majority of $\mu$Jy sources are at redshifts $z < 1.5$.
Simulations show that star forming galaxies should not be detectable for
$z > 3$ in the deepest radio images currently available without
evolution to include sources much more luminous than Arp 220 (Chapman
\etal\ 2002).

The $\mu$Jy sources with optically faint counterparts ($I \gapeq 24$)
are most likely to have $z > 2$ (Richards \etal\ 1999; Chapman \etal\
2003a).  The main evidence for dust in these is their high detectability
rate with SCUBA at 850\micron\ (e.g.\ Barger \etal\ 2000; Chapman \etal\
2003a).  In addition, the optical counterparts tend to be redder than
other field galaxies with the same $I$ magnitudes and become
progressively redder towards fainter magnitudes which also suggests the
presence of dust in the host (Chapman \etal\ 2003a).  Unlike the SCUBA
galaxies, $\mu$Jy radio galaxies usually have optical counterparts when
one looks hard enough (Richards \etal\ 1999).  Since these are rather
faint, redshifts are difficult to obtain in the optical.  Redshifts can
be crudely estimated using the ratio of SCUBA and radio fluxes (e.g.\
Carilli \&\ Yun 1999), however there are nasty degeneracies with dust
temperature to contend with (Blain 1999).

\section{Starburst Reddening and the sub-mm background}

Here I summarize the status of a recent debate within the field over the
last four or five years: which population of galaxies contributes the
most to the total star formation rate density at high redshift?  The two
top contenders are the optically selected LBG population and the
``SCUBA'' sub-mm and mm galaxies.  Table 2 summarizes some properties of
these contenders and their estimated contribution to the sub-mm
background $S_{\rm 850\mu{m}} = 44\, {\rm Jy\, deg^{-2}}$ (Fixsen \etal\
1998).

\begin{table}
\caption{What dominates the SFR at $z \sim 3$? }
\begin{center}
\begin{tabular}{lll}&& \\[-2mm] \hline \hline \\[-3mm]
Property & LBGs & SCUBA galaxies \\
&&\\[-3mm] \hline \\[-3mm]
Detectability & ``easy'' \&\ numerous         & difficult \&\ rare \\
 & $\sim 6000\zsp {\rm deg}^2$ to $R \sim 25.5$ & $\sim 300\zsp {\rm deg}^2$ to $\sim 6.5$mJy \\
 &                                            & confusion limited \\
 & (Steidel \etal\ 2003)                      & (e.g.\ Hughes \etal\ 1998) \\&&\\[-3mm] \hline \\[-3mm]
Luminosity & $L_{\rm UV} \sim 10^{10-11}\, L_\odot$ & $L_{\rm FIR} \sim 10^{12-13.5}\, L_\odot$ \\ 
           & (without dust correction)        & \\&&\\[-3mm] \hline \\[-3mm]
$L_{\rm FIR}/L_{\rm UV}$ & $\lapeq 100$ (average 5 -- 8) & 30 -- 3000$^a$ \\
\zsp local analog  & UV-bright starbursts & ULIRGs (Goldader \\
              & (MHC99)              & \etal\ 2002; AS00) \\&&\\[-3mm] \hline \\[-3mm]
Contribution to $S_{850}$ & 93\%\    & 100\%\ (by definition\ldots) \\
\zsp after corrections & $\sim$1.3 completeness & $\sim$5 completeness \\
                  & $\sim$6 dust absortion &  \\
                  & (AS00)           & Chapman \etal\ (2000) \\&&\\[-3mm] \hline \\[-3mm]
\end{tabular}
\end{center}
$^a$ One deviant sub-mm galaxy is SMMJ16358+4057 with 
$L_{\rm FIR}/L_{\rm UV} \sim 10$ (Smail \etal\ 2003). 
\end{table}

As noted above, LBGs strongly resemble UV bright starbursts.  One of the
strongest correlations seen in local UV bright starbursts is the
so-called IRX-$\beta$ relationship, shown in Fig.~1 (Meurer \etal\ 1995;
MHC99).  This relationship shows that the ratio of dust emission in the
FIR to (residual) UV emission (infra-red excess or IRX) correlates with
the UV spectral slope $\beta$ (defined by the spectrum $f_\lambda
\propto \lambda^\beta$).  Since IRX is basically a measure of dust
extinction this means that starburst redden as they become more
extincted.  The simplest explanation for this correlation is that the
dust has a strong diffuse foreground contribution to its distribution,
i.e.\ it behaves like a foreground screen (e.g.\ Witt \&\ Gordon 2000;
MHC99; Calzetti \etal\ 1994; Witt \etal\ 1992).  Indeed, correcting
survey data for dust absorption using simple reddening models greatly
improves the consistency of multi-wavelength SFRD$(z)$ plots (Calzetti
1999).  

It should be noted that not all local star forming galaxies obey this
relationship.  While it works well for the UV-bright calibrating sample
whose members have $L_{\rm bol} \lapeq 10^{11.5}$ it does not work well
for ULIRGs which typically have $\log(F_{\rm FIR}/F_{1600}) > 2$ and
fall well above the IRX-$\beta$ relationship (Goldader \etal\ 2002).  A
significant fraction ($\sim 30$\%) of normal (i.e.\ non starburst)
galaxies fall below the IRX-$\beta$ relationship, presumably due to
strong contamination from intermediate age populations in the near UV
(Seibert, 2003).

\begin{figure}
\plotone{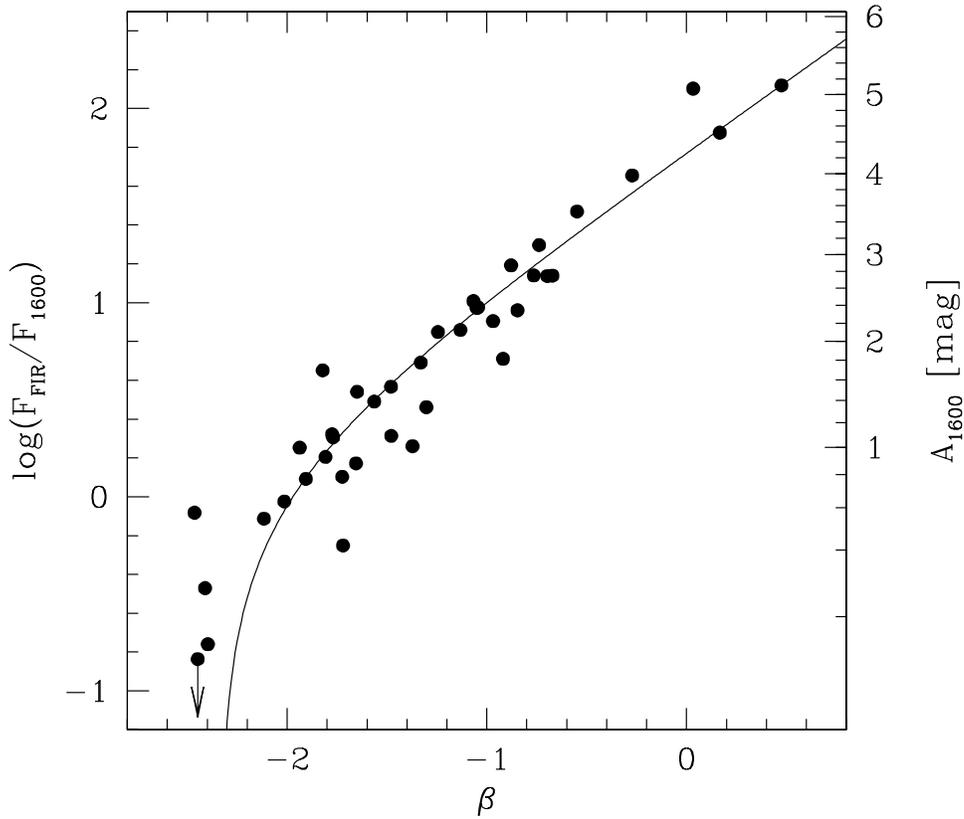}
\caption{The IRX-$\beta$ correlation (from MMC99). The UV spectral
slope $\beta$ is plotted on the abscissa.  On the ordinate is plotted
IRX, the ratio of fluxes in the FIR and and UV (at 1600\AA). At right
IRX is converted to the effective UV extinction $A_{1600}$.  The data
points represent the UV-bright starburst sample of MMC99, derived from
the IUE atlas of star forming galaxy (Kinney \etal\ 1993).  The curve is
a simple linear fit of $A_{1600}$ as a function of $\beta$. }
\end{figure}

Be that as it may, the IRX-$\beta$ correlation seems to work well for
strongly star forming galaxies with modest amounts of dust absorption.
While there is some distaste for the foreground screen geometry in the
literature (e.g.\ Charlot \&\ Fall 2000), we need not fixate on the
interpretation to use the correlation.  We have seen the myriad ways
that LBGs resemble local starbursts, so it seems reasonable to suppose
that they obey the same IRX-$\beta$ correlation.  If so, then we can
estimate the total extinction correction UV flux (and integrated cosmic
UV flux density), and hence star formation rate of LBGs from just their
rest-frame UV-flux and colors.  This was done by MHC99. An even better
job, using similar and other methods, was done by AS00.  Both papers
find that only 12\%\ -- 20\%\ of the UV flux emitted at rest $\lambda
\approx 1600$\AA\ reaches the earth (cf.\ Vijh \etal\ 2003, and
references therein).

It is hard to test whether the IRX-$\beta$ relation holds for LBGs,
since the predicted sub-mm fluxes are below the SCUBA confusion limit
(MHC99).  AS00 show that there are a few individual LBGs that have been
directly detected with SCUBA and these largely obey the IRX-$\beta$
relation.  One lensed LBG, MS1512+36-cB58, falls somewhat below the
IRX-$\beta$ relationship, albeit with large error bars (Baker \etal\
2001).  If a large fraction of LBGs have similar properties, then the
IRX-$\beta$ relationship would over predict their contribution to the
total SFRD at $z \sim 3$.

Fortunately, while LBGs in general are not individually
visible at other dust-insensitive wavelengths, stacked measurements of
fluxes in the radio and X-ray can be used to test whether LBGs have
similar dust extinction properties as nearby UV-bright starbursts.

The radio result is given as a note in proof to MHC99 and repeated in
Table~3.  Radio fluxes of the the ten $U$-dropout galaxies
in the Hubble Deep Field North (HDF-N) with the highest predicted 850\micron\
SCUBA fluxes were computed by assuming that the galaxies obey the local
FIR to radio correlation, have a radio spectral slope $\alpha =
-0.7$ ($f_\nu \propto \nu^\alpha$) and that the FIR flux is the
reprocessed UV flux.  The stacked radio fluxes, kindly provided by E.\
Richards, agree remarkably well with the predictions for these ten
sources.  AS00 also do an analysis of the stacked LBG fluxes at 20cm
using the Richards \etal\ radio map.  Their total observed and predicted
fluxes are $105 \pm 81\, \mu$Jy and $114 \pm 36\, \mu$Jy respectively,
similar to what is reported in MHC99 but with larger errors.  The
difference is in part due to AS00 stacking {\em all\/}
their 46 HDF LBGs for the analysis.  This includes many with
very little predicted flux which contribute the majority of the error
budget.  In addition, AS00 derive a higher error per
beam by integrating over the beam, which overestimates the error per
source.  Hence the stacked radio flux estimate is probably more
significant than implied by AS00.

\begin{table}
\caption{Sum of radio fluxes: HDF $U$-dropouts with the ten brightest
predicted 850\micron\ fluxes (MHC99). }
\begin{center}
\begin{tabular}{ccc}&& \\[-2mm] \hline \hline \\[-3mm]
Wavelength & \multicolumn{2}{c}{Flux density ($\mu$-Jy)} \\
(cm)       & predicted & observed \\ 
&&\\[-3mm] \hline \\[-3mm]
3.5        &  28 &  $27 \pm  5$ \\
20         & 100 & $105 \pm 24$ \\ &&\\[-3mm] \hline \\[-3mm]
\end{tabular}
\end{center}
\end{table}

X-rays can also be used to probe high-$z$ star formation.  The Chandra
soft X-ray band corresponds to 2-8 Kev at $z=3$, ``hard'' enough to pass
through the ISM of galaxies virtually unattenuated.  Seibert, Heckman
\&\ Meurer (2002) compared the stacked soft X-ray flux of HDF-N AGN free
$U$-dropouts given by Brandt \etal\ (2001) with predictions from dust
reddening models.  The results are consistent with a variety of
plausible dust reddening laws including the IRX-$\beta$ correlation from
MHC99, the Calzetti \etal\ (2000) starburst ``obscuration curve'', and
the homogeneous and clumpy foreground dust screen models from Witt \&\
Gordon (2000).  The results are not consistent with the rest-frame UV
flux from LBGs being like that of ULIRGs which have IRX values on the
order of 10$^2$ to 10$^{3.5}$ (Goldader \etal\ 2002).  In fact, if LBGs
had similar values they would be easily observable individually in
X-rays, the sub-mm, and radio (and they are not).  Similarly the
scenario that LBGs are not extincted at all under-predicts the stacked
X-ray flux by a factor of six.  Nandra \etal\ (2002) do an independent
analysis of stacked X-ray results in the HDF-N from Chandra and reach
similar conclusions.

The stacked radio and X-ray analyses are both consistent with the local
starburst reddening relation applying to high-$z$ LBGs.  If so, then
they would dominate total SFR density at $z \approx 3$ (AS00).  However,
there may be a fly in the ointment.  As earlier mentioned, in the MHC99
and AS00 picture we would expect the LBGs to have 850\micron\ fluxes up
to $\sim 1$ mJy.  While such fluxes are at or below the confusion limit
of blank field SCUBA observations it may be possible to detect such
sources through gravitational lensing.  Smail \etal\ (2002b) constrain
the faint counts at 850 \micron\ using SCUBA observations towards
lensing clusters.  A little over half of their sample are undetected in
their HST $I$ images, with the non-detections preferentially being at
the (sub-mm) faint end.  These faint end sources have magnification
corrected 850\micron\ micron fluxes like the predictions for LBGs.  The
implication is that the LBGs are not showing up in SCUBA observations
presumably because the local IRX-$\beta$ relationship overpredicts
their 850\micron\ flux as it does for MS1512+36-cB58. Instead the
850\micron\ background is actually dominated by star forming galaxies
like local ULIRGs (Goldader \etal\ 2002) - totally hidden by dust, as is
also the case with the brightest SCUBA sources.

While this scenario may be correct, a careful examination of Smail
\etal\ (2002b) suggests that it is built on a shaky foundation.  Their
claim to having resolved the 850\micron\ background rests on four faint
($S/N = 3$ to 5) 850\micron\ detections.  These have only have lower
limits to their magnifications (presumably because their positions are
uncertain, since they have no optical, NIR, or radio counterparts).
Their Monte-Carlo simulations indicate that at least one of these
sources should have a source plane 850\micron\ flux in the 0.5 to 1 Jy
range, the only sources in their survey that could be that faint (or
fainter).  However, using the lower limit magnifications shows that none
of the source plane fluxes need actually be fainter than 1.6 Jy, while
the brightest predicted 850\micron\ flux for the LBGs in the HDF-N is
1.8 mJy (MHC99).  Smail \etal\ are not convincingly dipping deep into
the expected realm of LBGs with their SCUBA observations.  The lack of
optical counterparts at the faint end also does not thoroughly rule out
the presence of LBGs.  They state a detection limit of $I \sim 26$ in
their HST images which corresponds to a typical brightness seen in LBGs.
However, this is the detection limit at $S/N = 2$ for a point source.
To do this well the detection limit should be stated at a higher $S/N$
(at least 3) and be calculated for typical LBG sizes (corrected for
lensing in their case).  I expect this would lower the limiting mag to
$I \sim 25$ or brighter. Figure 15 of AS00 implies that the optical
counterparts of the $\sim 1-2$ Jy sources have magnitudes over a wide
range $I \sim 24$ to 27 ABmag in the source plane.  Smail \etal\ have
not convincingly ruled this out yet.

I conclude that the nature of the sources dominating the sub-mm
background is not yet well determined.  It is clear that the leading
contenders are galaxies at $z \gapeq 2$ whose bolometric output is
dominated by dust, and that these galaxies dominate the star formation
rate density at these redshifts. It is not yet settled whether these
galaxies are detectable in the rest-frame UV.  Additional observations
of lensing clusters in the mm and sub-mm would improve the statistics on
the 850\micron\ counts at and below 2 mJy, while deeper optical
observations of the faintest SCUBA sources (e.g.\ with ACS on HST) are
needed to make a fair and convincing test of the LBG scenario.

\section{What the future holds}

High redshift observational cosmology is one of the most active fields
of astronomical research, as can be seen by sifting through the daily
offerings on astro-ph.  Progress is rapid and happening on many fronts.  

The installation of the Advanced Camera for Surveys on HST in 2002
allows routine optical imagining to Hubble Deep Field depths in fields
with twice the area and angular resolution of WFPC2 (Ford \etal\ 2002).
The first ACS results from the GOODS project (Giavalisco \etal\ 2003)
are soon to come out in a special ApJ Letters volume.  The results
include reported color evolution in the LBG population, suggesting less
dust is present in the $z \sim 4$ population compared to the $z \sim 3$
population (Idzi \etal\ 2003; Papovich \etal\ 2003).

The recent successful launches of the
GALEX\footnote{http://www.galex.caltech.edu/} and
SIRTF\footnote{http://sirtf.caltech.edu/} satellites will allow
excellent survey capabilities in the vacuum UV (0.13 -- 0.3 \micron) and
infrared (3.5 -- 160 \micron) respectively.  Together they will provide
the definitive local calibration of the IRX-$\beta$ relationship.  GALEX
will extend the redshift range for LBG selection to $z < 2$.  The SIRTF
observations of the GOODS project will directly detect normal galaxies
and LBGs out to $z \sim 5$ LBGs at 3.5 -- 8 \micron (rest frame NIR)
which allows probing of their evolved stellar populations.  At 24
\micron\ they should also be detect rest frame $\sim$ 7 \micron\
reprocessed PAH emission from galaxies with $L_{\rm FIR} \gapeq 10^{11}$
out to $z \sim 2$.

Recent progress in sub-mm galaxy research includes an improved
efficiency in obtaining redshifts.  Chapman \etal\ (2003b) present
optical redshifts for 10 sub-mm sources also detected in the radio.
Soon that group will publish an expanded sample of about 70 optical
spectroscopic redshifts of sources detected at both radio and sub-mm
wavelengths. They find that sub-mm galaxies mostly do indeed have the
high redshift ($z \sim 2.5$) that was expected.  This work provides the
first accurate redshift distribution and space density measurements of
the sub-mm population.  While more progress could be made in identifying
the faintest SCUBA sources (as outlined in \S~3), what is really needed
is more and reliable detections of the $\lapeq 1$ mJy population,
i.e. somewhat {\em below\/} the confusion limit.  It seems unlikely that
there will be enough clusters not already looked at by the Smail, Blain,
Ivison group and other SCUBA researchers to add substantial numbers of
faint sources.

Unfortunately, at the longest wavelengths (160 \micron), the SIRTF PSF
has FWHM = 38\as\ and hence the detections will also be confusion
limited at around 7 to 19 mJy (Xu \etal, 2001), and thus SIRTF is not
likely to settle the debate on the source composition of the sub-mm/FIR
background.  SOFIA will provide higher resolution imaging ($\sim 8''$ at
200\micron) but due to the high sky background will generally be limited
to galaxies with $z <
1$\footnote{http://sofia.arc.nasa.gov/Science/science/sci\_opport\_galaxies.html}.
Bolometer arrays larger than SCUBA, such as
SCUBA-II\footnote{http://www.jach.hawaii.edu/JACpublic/JCMT/JCMT\_developments/SCUBA2/scuba2.html}
also will not address this issue since it is angular resolution, and
hence telescope aperture that limits us from resolving the FIR - mm
background.  Existing or soon to be completed sub-mm to mm arrays such
as the SMA\footnote{http://sma-www.harvard.edu/} will help to pin down
the position of the brightest SCUBA sources.  However, to truly resolve
the sub-mm to mm background we must await significant completion of ALMA
-- the Atacama Large Millimeter
Array\footnote{http://www.alma.nrao.edu/}.  When completed in 2012 it
will be able to reach flux densities of $\sim 0.1$ mJy (well into the
expected fluxes of LBGs) in about half an hour at 850 \micron. With a
resolution of 0.1\as\ or better (depending on array configuration) the
observations should be well out of the confusion limit.  Combining ALMA
data with (then ancient archival) data from HST and hopefully new data
from JWST (the James Webb Space Telescope) will allow a direct estimate
of the bolometric output, and hence star formation rate, of all types of
high-$z$ galaxies.  This will make the question of whether galaxies obey
the IRX-$\beta$ relationship somewhat moot.

\section{Summary}

The papers reviewed here have shown strong evidence that almost all
types of high-$z$ sources contain dust.  The only case where the
evidence is not yet convincing is the quasar absorption line systems,
particularly at the column densities of the Lyman forest.  In that case,
it is not even clear that we are dealing with galaxies.  All other cases
involve galaxies with at least some inferred star formation.  Hence at
high redshift, as in the local universe, star formation and dust are
correlated.  The dominant location of high redshift star formation
remains debatable.  While it remains plausible that the majority of star
formation at $z > 2$ remains completely invisible shortwards of the
rest-frame FIR, there is a strong body of evidence that moderately dust
obscured but rest-frame UV-bright galaxies should dominate the star
formation in the early universe.  It is likely that this issue will not be
definitively settled until the sub-mm background is fully resolved. 

\acknowledgements I thank Adolf Witt for cajoling me to become Daniela
Calzetti's replacement at the conference.  My talk and this review
benefited from discussions and correspondence with John Blakeslee,
Daniela Calzetti, Mark Dickinson, Tim Heckman, Jason Prochaska, and
Chuck Steidel.  I thank the anonymous referee and Andrew Blain for
suggestions that have improved the paper, allowing me to go deeper into
the issues than I did at Estes Park.


\begin{references}
\reference Adelberger, K.L. \&\ Steidel, C.C. 2000, \apj, 544, 218 (AS00)
\reference Adelberger, K.L.,  Steidel, C.C., Shapley, A.E., \&\ Pettini,
M. 2003, \apj, 584, 45 
\reference Aguirre, A.N. 1999, \apj, 512, L19
\reference Aguirre, A., \&\ Haiman, Z. 2000, \apj, 532, 28
\reference Baker, A.J., Lutz, D., Genzel, R., Tacconi, L.J., \&\
Lehnert, M.D. 2001, \aap, 372, 37
\reference Barger, A.J., Cowie, L.L., \& Richards, E.A., 2000, \aj, 119, 2092
\reference Blain, A.W. 1999, \mnras, 309, 955
\reference Blain, A.W., Smail, I., Ivison, R.J., Kneib, J.-P. \& Frayer,
D.T. 2002, Physics Reports, 369, 111
\reference Blakeslee, J.P., \etal\ 2003, \apj, 589, 693
\reference Brandt, W.N., Hornschemeier, A.E., Schnedier, D., Alexander,
D.M., Bauer, F.E., Garmire, G.P., \&\ Vignali, C. 2001, \apj, 558, L5
\reference Calzetti, D. \&\ Heckman, T. 1999, \apj, 519, 27
\reference Calzetti, D., Kinney, A.L., \&\ Storchi-Bergmann, T. 1994,
\apj, 429, 582
\reference Calzetti, D. 1999, in ``Building the Galaxies: from the
Primordial Universe to the Present (XXXIVeme Recontres de Moriond)'',
(Editions Frontieres, Paris), Eds. F. Hammer, T.X. Thuan, V. Cayatte,
B. Guiderdoni and J.T. Thanh Van, (astro-ph/9907025)
\reference Calzetti, D., Armus, L., Bohlin, 
R.C., Kinney, A.L., Korneef, J., \&\ Storchi-Bergmann, T. 2000, \apj,
533, 682
\reference Calzetti, D. 2001, \pasp, 113, 1449
\reference Carilli, C.L., \&\ Yun, M.S. 1999, \apj, 513, L13
\reference Chapman, S.C., \etal\ 2000, \mnras, 319, 318
\reference Chapman, S.C., Lewis, G.F., Scott, D., Borys, C., \&\
Richards, E. 2002, \apj, 570, 557
\reference Chapman, S.C., Barger, A.J., Cowie, L.L, Scott, D., Borys,
C., Capake, P., Fomalont, E.B., Lewis, G.F., Steffen, A.T., Wilson, G.,
\&\ Yun, M. 2003a, \apj, 585, 57
\reference Chapman, S.C., Blain, A.W., Ivison, R.J, \&\ Smail,
I.R. 2003, Nature, 422, 6933
\reference Charlot, S., \&\ Fall, S.M. 2000, \apj, 539, 718
\reference Dannerbauer, H., Lehnert, M.D., Lutz, D., Tacconi, L.,
Bertoldi, F., Carilli, C., Genzel, R., \&\ Menten, K. 2002, \apj, 573,
473
\reference de Jong, T., Klein, U., Wielebenski, \&\ Wunderlich, E. 1985,
\aap, 147, L6
\reference Ellison, S.L., Yan, L., Hook, I.M., Pettini, M., \&\ Shaver,
P. 2001, \aap, 379, 393
\reference Elston, R., Rieke, G.H., \&\ Rieke, M.J. 1988, \apj, 331, L77
\reference Fall, S.M., \&\ Pei, Y.C. 1993, \apj, 402, 479
\reference Fan, X. \etal\ 2001, \aj, 122, 2833
\reference Fan, X., \etal\ 2003, \aj, 125, 1649
\reference Fixsen, D.J., Dwek, E., Mather, J.C., Bennett, C.L., \&\ Shafer,
R.A. 1998, \apj, 508, 123
\reference Fomalont, E.B., Kellermann, K.I., Partridge, R.B., Windhorst,
R.A., \&\ Richards, E.A. 2002, \aj, 123, 2402
\reference Ford, H.C., \etal\ 2002, Proc. SPIE, 4854, 81
\reference Frayer, D.T., Armus, L., Scoville, N.Z., Blaian, A.W.,
Reddy, N.A., Ivison, R.J. \&\ Smail, I. 2003, \aj, 126, 73
\reference Giavalisco, M., Steidel, C.C., \&\ Macchetto, F.D. 1996,
\apj, 470, 189 
\reference Giavalisco, M, \etal\ 2003, ApJL, in press (astro-ph/0309105)
\reference Goldader, J.D., Meurer, G.R., Heckman, T.M., Seibert, M.,
Sanders, D.B., Calzetti, D., \&\ Steidel, C.C. 2002, \apj, 568, 651
\reference Goobar, A., Bergstr\"{o}m, L., M\"{o}rtsell, E. 2002, \aap,
384, 1
\reference Gordon, K.D., Calzetti, D., \&\ Witt, A.N. 1997, \apj,
487, 625
\reference Guiderdoni, B., Bouchet, F.R., Puget, J.L., Lagache, G., \&\
Hivon, H. 1997, Nature, 390, 257
\reference Helou, G., Soifer, B.T., \&\ Rowan-Robinson, M. 1985,
\apjl, 298, L7
\reference Hughes, D., et al. 1998, Nature, 394, 241
\reference Idzi, R., Somerville, R., Papovich, C., Ferguson, H. C.,
Giavalisco, M., Kretchmer, C. \&\ Lotz, J., 2003, ApJL, in press
(astro-ph/0308541) 
\reference Irwin, M.J., \&\ McMahon, R.G., 1991, PASAu, 9, 246
\reference Ivison, R.J. \etal\ 2002, \mnras, 337, 1
\reference Kinney, A.L., Bohlin, R.C., Calzetti, D., Panagia, N., \&\
Wyse, R.F.G. 1993, \apjs, 86, 5
\reference Kodaira, K., \etal\ 2003, PASJ, 55, L17
\reference Kunth, D., Mas-Hesse, J.M., Terlevich, E., Terlevich, R.,
Lequeux, J., \&\ Fall, S.M. 1998, \aap, 334, 11
\reference Lisenfeld, U., V\"olk, H.J., \&\ Xu, C.  1996, \aap, 306, 677
\reference Lowenthal, J.D., Koo, D.C., Guzm\'an, R., Gallego, J.,
Phillips, A.C., Faber, S.M., Vogt, N.P., Illingworth, G.D., \&\
Gronwall, C.  1997, \apj, 481, 673
\reference Madau, P., Ferguson, H.C., Dickinson, M.E., Giavalisco,
M., Steidel, C.C., \&\ Fruchter, A. 1996, \mnras, 283, 1388
\reference Madau, P., Pozzetti, L., \&\ Dickinson, M. 1998, \apj, 498,
106
\reference Meurer, G.R., Heckman, T.M., Leitherer, C., Kinney, A.,
Robert, C., \&\ Garnett D.R. 1995, \aj, 110, 2665
\reference Meurer, G.R., Heckman, T.M., Lehnert, M.D., Leitherer,
C., \&\ Lowenthal, J.  1997, \aj, 114, 54
\reference Meurer, G.R., Heckman, T.M., \& Calzetti, D. 1999, \apj,
521, 64 (MHC99)
\reference Nandra, K., Mushotzky, R.F., Arnaud, K., Steidel, C.C.,
Adelberger, K.L., Gardner, J.P, Teplitz, H.I., \&\ Windhorst, R.A. 2002,
\apj, 576, 625
\reference Omont, A., Petitjean, P., Guilloteau, S., McMahon, R.G.,
Solomon, P.M., \&\ P\'econtal, E. 1996, Nature, 382, 428
\reference Paerels, F, Petric. A., Telis, G., \&\ Helfand, D.J., 2002,
\baas, 201, 97.03
\reference Papovich, C., Dickinson, M., \&\ Ferguson, H.C. 2001, \apj,
559, 620
\reference Papovich, C., \etal\ 2003, ApJL, submitted
\reference Partridge, R.B., \&\ Peebles, P.J.E. 1967, \apj, 147, 868
\reference Peacock, J.A., \etal\ 2000, \mnras, 318, 535
\reference Pettini, M., Kellogg, M., Steidel, C.C., Dickinson, M.,
Adelberger, K.L., \&\ Giavalisco, M. 1998, \apj, 508, 539
\reference Pettini, M., Steidel, C.C., Adelberger, K.L., Dickinson, M.,
\&\ Giavalisco, M. 2000, \apj, 528, 96
\reference Pettini, M., Madau, P., Bolte, M., Prochaska, J.X., Ellison,
S.L., \&\ Fan, X. 2003, ApJ (accepted, astro-ph/0305413)
\reference Pritchet, C.J. 1994, \pasp, 106, 1052
\reference Prochaska, J.X., Howk, J.C. \&\ Wolfe, A.M. 2003, Nature,
423, 57
\reference Richards, E.A., Fomalont, E.B., Kellermann, K.I., 
Windhorst, R.A., \&\ Partridge, R.B., Cowie, L.L., \&\ Barger,
A.J. 1999, \apj, 526, L73
\reference Rhoads, J.E., Malhotra, S., Dey, A., Stern, D., \&\ 
Spinrad, H. 2000, \apj, 545, L85
\reference Rhoads, J.E., Dey, A., Malhotra, S., Stern, D., Spinrad, H.,
Januzzi, B., Dawson, S., Brown, M.J., \&\ Landes, E. 2003, \apj, 125,
1006 
\reference Seibert, M., Heckman, T.M., \&\ Meurer, G.R. 2002, \aj, 124,
46 
\reference Seibert, M. 2003, Ph.D. Thesis, The Johns Hopkins University
\reference Shapley, A.E., Steidel, C.C., Pettini, M., \&\ Adelberger,
K.L. 2003, \apj, 588, 65
\reference Smail, I., Owen, F.N., Morrison, G.E., Keel, W.C., Ivison,
R.J., \&\ Ledlow, M.J. 2002a, \apj, 581, 844
\reference Smail, I., Ivison, R.J., Blain, A.W., and Kneib, J.-P. 2002b,
\mnras, 331, 495
\reference Smail, I., Chapman, S.C., Ivison, R.J., Blain, A.W., Takata,
T., Heckman, T.M., Dunlop, J.S., \&\ Sekiguchi, K. 2003, \mnras, 342, 1185
\reference Songaila, A. 2001, \apj, 561, L153
\reference Steidel, C.C., Giavalisco, M., Pettini, M., Dickinson,
\&\ M., Adelberger, K.L. 1996a, \apjl, 462, L17
\reference Steidel, C.C., Giavalisco, M., Dickinson, M., \&\ Adelberger,
K.L. 1996b, \aj, 112, 352
\reference Steidel, C.C., Adelberger, K.L., Shapley, A.E., Pettini, M. 
Dickinson, M., \&\ Giavalisco, M. 2003, 592, 728
\reference Totani, T., Yoshii, Y., Fumihide, I., Toshinori, M., \&\
Motohara, K. 2001, \apj, 558, L87.
\reference Tremonti, C.A., Calzetti, D., Leitherer, C., Heckman, T.M. 
2001, \apj, 555, 322
\reference Vijh, U.P., Witt, A.N. \&\ Gordon, K.D., 2003, \apj, 587,
533
\reference Witt, A.N. \&\ Gordon,  K.D. 2000, ApJ, 528, 799
\reference Witt, A.N., Thronson, H.A., \&\ Capuano J.M.  1992, \apj, 393, 611
\reference Xu, C., Lonsdale, C., Shupe, D.L, O'Linger, J., \&\ Masci,
F. 2001, \apj, 562, 179
\end{references}
\end{document}